\documentclass[aps,prl,reprint,superscriptaddress]{revtex4-1}

\usepackage[T1]{fontenc}
\usepackage[utf8]{inputenc}
\usepackage{gensymb}
\usepackage{graphicx}

\begin{document}


\title{Coalescence and anti-coalescence of surface plasmons on a lossy beamsplitter}


\author{Benjamin Vest}
\email[e-mail :]{benjamin.vest@institutoptique.fr}

\author{Marie-Christine Dheur}
\affiliation{Laboratoire Charles Fabry, Institut d'Optique, CNRS, Université Paris-Saclay, 91127 Palaiseau cedex, France.}

\author{Eloïse Devaux}
\author{Thomas W. Ebbesen}

\affiliation{Institut de Science et d'Ingénierie Supramoléculaires, CNRS, Université de Strasbourg, 67000 Strasbourg, France.}

\author{Alexandre Baron}

\affiliation{Centre de Recherche Paul Pascal, CNRS, 33600 Pessac, France.}

\author{Emmanuel Rousseau}
\affiliation{Laboratoire Charles Coulomb, UMR CNRS-UM 5221, Université de Montpellier, 34095 Montpellier, France.}

\author{Jean-Paul Hugonin}

\affiliation{Laboratoire Charles Fabry, Institut d'Optique, CNRS, Université Paris-Saclay, 91127 Palaiseau cedex, France.}

\author{Jean-Jacques Greffet}
\author{Gaétan Messin}
\author{François Marquier}

\email[e-mail :]{francois.marquier@institutoptique.fr}

\affiliation{Laboratoire Charles Fabry, Institut d'Optique, CNRS, Université Paris-Saclay, 91127 Palaiseau cedex, France.}

\date{\today}

%
%
%


\maketitle


\textbf{Surface plasma waves are collective oscillations of electrons that propagate along a metal-dielectric interface \cite{Ritchie:57}. In the last ten years, several groups have reproduced fundamental quantum optics experiments with surface plasmons. Observation of single-plasmon states \cite{Akimov:07,Cai:14}, wave-particle duality \cite{Kolesov:09,Dheur:16}, preservation of entanglement of photons in plasmon-assisted transmission \cite{Altewischer:02,Fasel:05,Ren:06}, and more recently, two-plasmon interference have been reported\cite{Heeres:13,Fakonas:14,DiMartino:14,Cai:14, wang2016}. While losses are detrimental for the observation of squeezed states, they can be seen as a new degree of freedom in the design of plasmonic devices, thus revealing new quantum interference scenarios. Here we report the observation of two-plasmon quantum interference between two freely-propagating, non-guided SPPs interfering on lossy plasmonic beamsplitters. As discussed in Refs. \cite{Barnett:98,Jeffers:00}, the presence of losses (scattering or absorption) relaxes constraints on the reflection and transmission factors of the beamsplitter, allowing the control of their relative phase. By using this degree of freedom, we are able to observe either coalescence or anticoalescence of identical plasmons.}

Two-particle interference, as a fundamental quantum feature, has been extensively studied with photons through the Hong-Ou-Mandel \cite{Hong:87} dip and has been recently observed with guided plasmons in a large variety of plasmonic circuits. It showed the possibility to generate pairs of indistinguishable single plasmons (SPPs), which is an important requirement for potential quantum information applications \cite{Knill:01,Chang:06,wang2016}. Despite the presence of losses, these experiments have shown that quantum effects remain observable. In these experiments, the propagation paths were lossy, but the beamsplitters were non-lossy. The presence of losses on the beamsplitter was studied in Refs \cite{Barnett:98,Jeffers:00}, where novel effects were predicted, including coherent absorption of single photon and N00N states \cite{Roger:15,Roger:16}. In our work, we designed several plasmonic beamsplitters with different sets of reflection and transmission factors, that were used in a plasmonic version of the Hong-Ou-Mandel experiment. Depending on the samples, coincidences detection measurements lead either to a HOM-like dip, i.e. a signature of plasmon coalescence, or a HOM-peak, that we associate to plasmon anti-coalescence.

\begin{figure}

\includegraphics[width=0.45\textwidth]{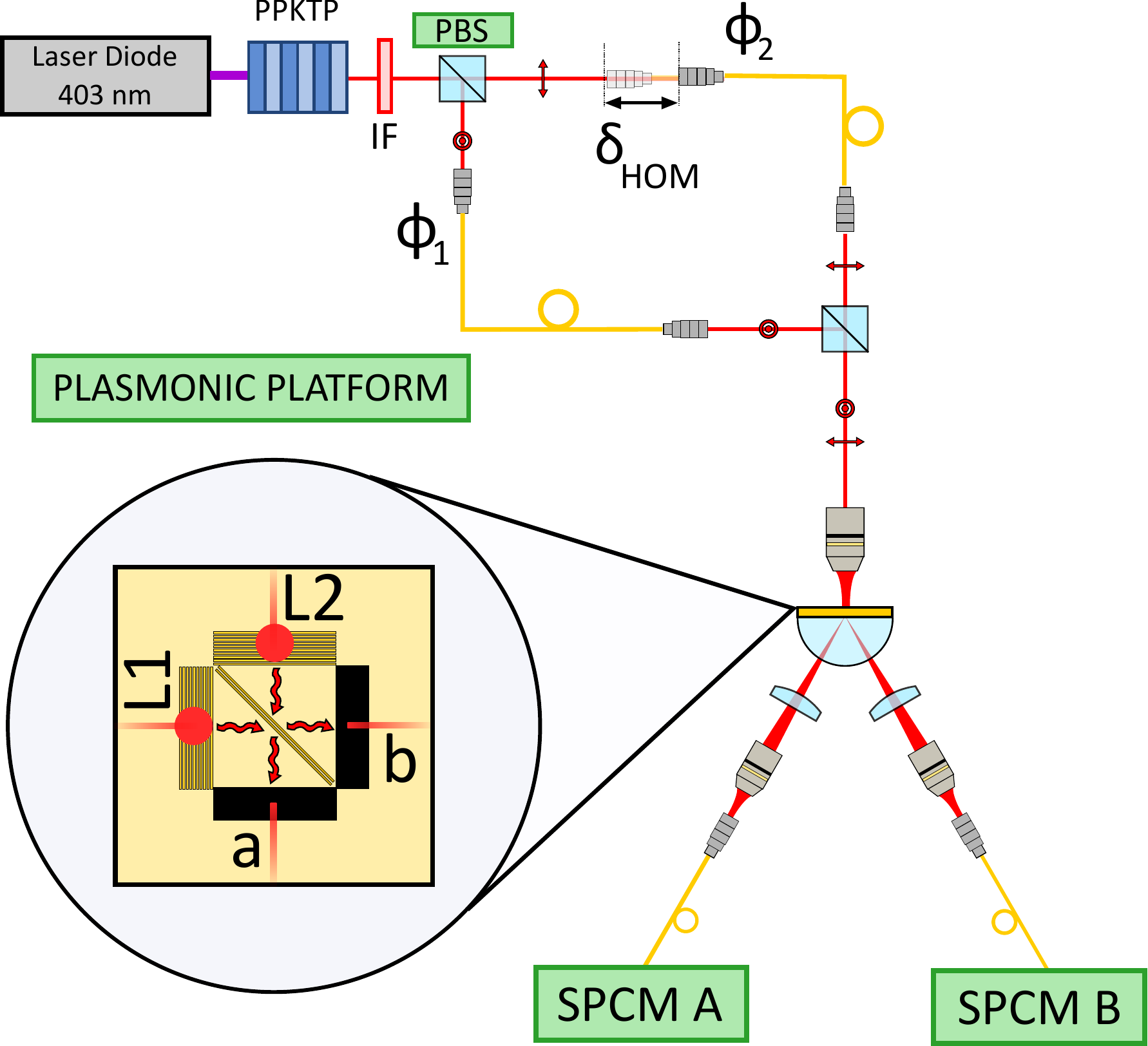}%
\caption{\label{fig1} Skecth of the setup. A PPKTP crystal is pumped by a laser diode at 403 nm and delivers pairs of orthogonally polarized photons at 806 nm. An interference filter (IF) removes the remaining pump photons. The near-infrared photons are separated by a polarizing beamsplitter (PBS), and excite the photonic modes $\phi_1$ or $\phi_2$, that are respectively converted by the SPP launchers L1 and L2 into plasmonic modes on a plasmonic platform. A delay $\delta_{\mathrm{HOM}}$ between the two SPPs can be settled changing the optical path of one of the photons after the PBS. The two single SPPs are recombined on a plasmonic beamsplitter and finally out-coupled to photonic modes $a$ and $b$. SPCMs A and B record detection counts respectively from output modes $a$ and $b$, and measure coincidences between the detectors.}
\end{figure}

Let first begin with a brief description of the experimental setup. It is based on a source of photon pairs. The photons of a given pair are sent to two photon-to-SPP converters, located at the surface of a plasmonic test platform. It has been shown recently that the photon number statistics are conserved when coupling the photonic modes to a plasmonic mode on such a device \cite{Tame:08}, so that incident single photons are converted into two single SPPs. These SPPs freely propagate on the metallic surface towards the two input arms of a plasmonic beamsplitter. Finally, the SPPs that reach the output of the platform are converted back to photons to be detected by single photon counting modules (SPCMs). A more detailed picture of the setup is depicted on Fig. \ref{fig1}. The single photons are generated by spontaneous parametric down conversion (SPDC) in a periodically-poled KTP crystal, pumped by a laser diode at 403 nm. This SPDC source delivers frequency degenerate pairs of orthogonally polarized photons at 806 nm. The photons are then separated by a polarizing beamsplitter (PBS), and injected in monomode fibers. Afterwards, they are sent separately on both input ports of the plasmonic platform. One of the fiber input collimator is placed on a motorized translation stage, allowing the displacement of the fiber's input along the optical axis. We can thus adjust the delay $\delta_{\mathrm{HOM}}$ between the SPPs's paths in order to observe interferences.

\begin{figure}
\includegraphics[width=0.45\textwidth]{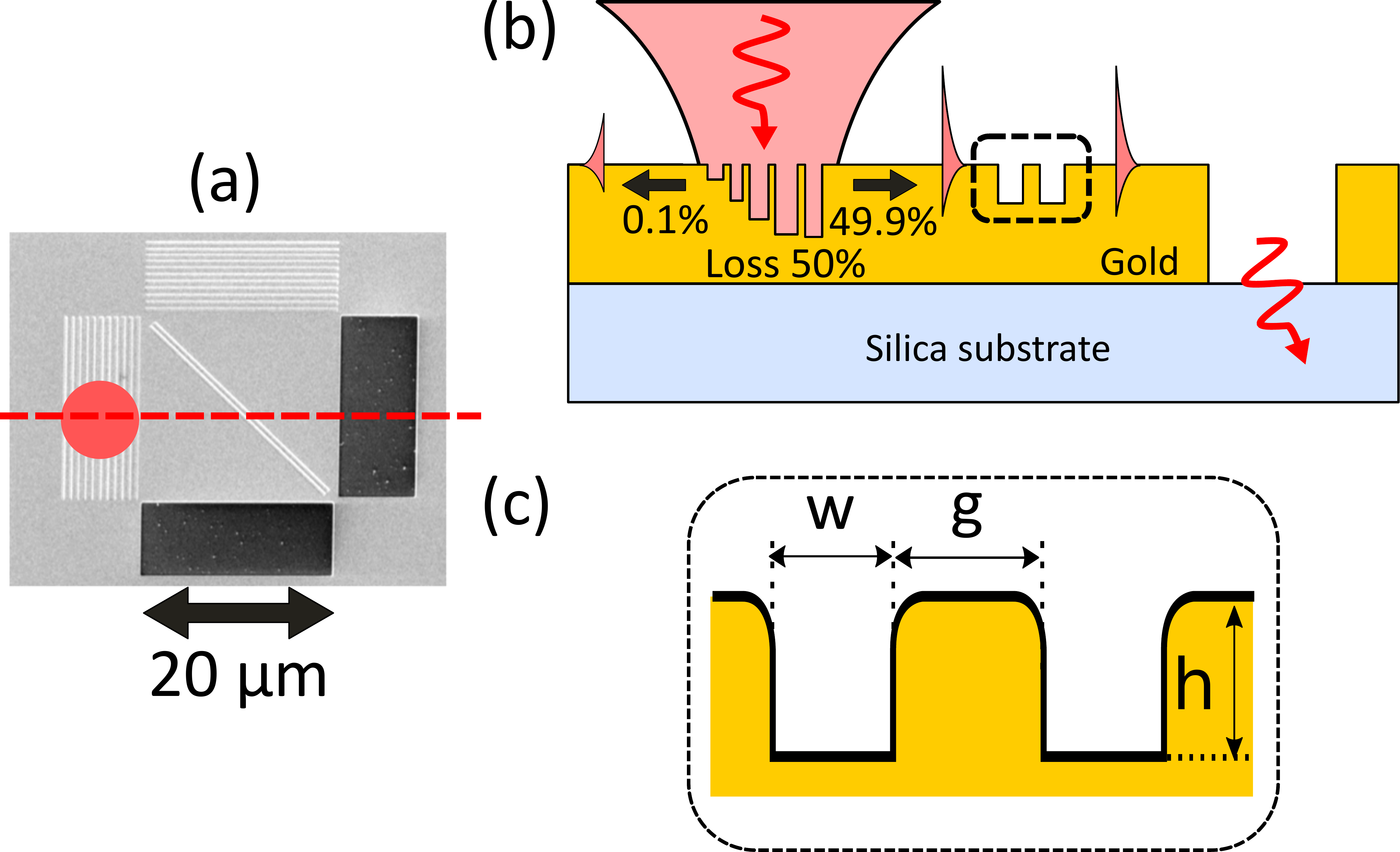}%
\caption{\label{fig2}(a) SEM picture of a plasmonic platform. The dotted red line represents the direction along which the section is depicted in (b). The red spot represents an incident gaussian beam on the SPP launcher. (b) Sectional drawing of the device. On the left, the first structure is a photon-to-SPP coupler. When single photons reach the grating, single SPPs are launched unidirectionaly toward the plasmonic beamsplitter (SPBS) (grooved doublet). They can be either reflected, transmitted or absorbed by the SPBS. The remaining SPPs propagate to the large outcoupling slits. With an efficiency of about 50\%, SPPs are converted back to photons in the silica substrate. (c) Close-up look on the SPBS. Dimensions of the SPBS are defined by three parameters, the grooves width $w$, the metal gap between the grooves $g$ and the height of the groove $h$.}
\end{figure}

The plasmonic platform consists in several elements that are etched on a 300 nm-thick gold film on top of a silica substrate, on a total 40x40 $\mu $m footprint (see Fig. \ref{fig2}). The input channels of the plasmonic platform are made of two unidirectional launchers (denoted as L1 and L2). Those asymmetric 11-grooves gratings have been designed to efficiently couple a normally incident Gaussian mode into directional SPPs \cite{Baron:11}.
The SPPs generated by each launcher then freely propagate and recombine on the surface plasmon beamsplitter (SPBS). It is made of two identical grooves in the metallic surface (see Fig. \ref{fig2}(c)), oriented at $45\degree$ with respect to the propagation direction of waves launched by L1 and L2. The succession of metal and air allows a scattering process that generates both a transmitted and a reflected SPP \cite{Gonzalez:06}. The complex reflection and transmission factors $r=|r| \exp(i \phi_{r})$ and $t=|t| \exp(i \phi_{t})$ of the SPBS are functions of the geometrical parameters of the SPBS (see Fig. \ref{fig2}(c)). By controlling the widths $w$, $g$, and the depth $h$ of these grooves, we can control the phase difference $\phi_{rt}=\phi_t-\phi_r$ . This phase control affects significantly the interferences, as we will see later. The SPPs then propagate towards two large out-coupling strip slits. They are decoupled into photons propagating in the glass substrate on the rear of the platform. The light is transmitted from the substrate to the free space by a hemispherical lens before being collected by two 75 mm-focal-length lenses at both output ports of the platform, and is injected by two focusing objectives in multimode fibers. Finally, two single-photon counting modules (SPCMs) detect the photon signal and gives a count rate of each output channel of the setup. Simultaneous operation of the two SPCMs enables the detection of coincidences of photons emerging from the two outputs of the SPBS.

For a lossless balanced beamsplitter, energy conservation and unitarity transformation of modes at the interface imposes  $t=\pm ir$ and $|t|=|r|=1 / \sqrt{2}$, so that the phase difference between $r$ and $t$ is $\phi_{rt} = \pm 90\degree$. When placed at the output of a Mach-Zehnder interferometer, the two outputs of the beamsplitter deliver a sinusoidal interference signal that displays a phase-shift $2\phi_{rt} = \pm 180\degree$. It follows that a maximum on a channel corresponds to a minimum on the other channel as expected from energy conservation arguments. The situation is however quite different in our experiment. In our case, a single SPP is transmitted with probability $|t|^2$, reflected with a probability $|r|^2$, but can be absorbed or scattered with a probability $1-|r|^2-|t|^2$. For a balanced SPBS \textit{in presence of losses}, $r$ and $t$ are constrained by the following inequality:

\begin{equation}
|t - r |^2 \leq 1,
\end{equation}
where the equality holds only if there are no losses. The previous relation releases all constraints on $2\phi_{rt}$. In other words, losses can here be considered as a new degree of freedom. It is therefore possible to design several beamsplitters where the amplitude of $r$ and $t$ and the relative phase $\phi_{rt}$ can be modified. As a direct consequence, interference fringes from both outputs of the BS can be found experiencing an arbitrary phase shift. 

Controlling those properties of the SPBS strongly affects the detection of events by the two SPCMs. It has been shown \cite{Barnett:98} that the coincidence detection probability, i.e. the probability for one particle pair to have its two particles emerging from separate outputs of the beamsplitter can be expressed as:

\begin{equation}
P(1_a,1_b) = |t|^4+ |r|^4 +  2 \Re(t^2 r^2) I,
\end{equation}

where $2 \Re(t^2 r^2) = t^2 r^{*2} +r^2 t^{*2} $, $a$ and $b$ label the output ports of the beamsplitter, and $I$ is an overlap integral between the two particles wavepackets. For non-overlapping wavepackets, $I=0$ and the previous relation reduces to:

\begin{equation}
P_{cl}(1_a,1_b) = |t|^4+ |r|^4.
\end{equation}

The particles impinging on the SPBS behave like two independent classical particles, as indicated by the subscript $cl$. For an optimal overlap between the particles ($I=1$), the coincidence probability can be written:

\begin{equation}
P_{qu}(1_a,1_b) = |t^2+r^2|^2 = P_{cl}(1_a,1_b) + 2 \Re(t^2 r^2),
\end{equation}

where the subscript $qu$ denotes the presence of the quantum interference term $2 \Re(t^2 r^2)$.

We now consider two cases. If $r=\pm it$, the probability $P_{qu}$ is zero. This is the same antibunching result that is obtained for a non lossy beam splitter \cite{Barnett:98}. This is the so-called Hong-Ou-Mandel dip in the correlation function. If we now consider $r=\pm t$ and $|r|=|t|= \frac{1}{2}$ we get $P_{qu}(1_a,1_b) = 2 P_{cl}(1_a,1_b)$. Here, we expect a peak in the correlation function.

The plasmonic chips were designed by solving the electrodynamics equations with a in-house code based on the aperiodic Fourier modal method \cite{Silberstein:01}. We designed two samples denoted as samples I and II. They correspond to the previous configurations $r=\pm it$ and $r=\pm t$ with $|r|=|t|= \frac{1}{2}$ respectively. The dimensions of each beamsplitter are reported in Table 1. The samples were characterized by splitting a 806nm-CW-laser beam and sending it on both input arms of the chip and recording the interference fringes at both output ports of the setup when increasing the relative delay $\delta_{\mathrm{HOM}}$. We then measured the average phase difference between the two signals recorded on the two output channels in order to get $\phi_{rt}$.

 \begin{table}
 \caption{\label{table} Dimensions of the plasmonic platform samples under study, as measured by a scanning electron microscope (width $w$ and metal gap $g$) and atomic force microscope (groove depth $h$). Notations refer to Figure \ref{fig2}(c). The fourth line reports expected values for the reflection and transmission factors $r$ and $t$ based on the numerical simulations of the target design. The last line reports estimations of the relative phase between the reflexion and transmission coefficients after characterization. Numbers between parenthesis are the target dimensions and relative phase of the devices as designed by numerical simulations.}
 \begin{ruledtabular}
\begin{tabular}{c|c|c|}

• & Sample I & Sample II \\ 
\hline 
w [nm] & 171 (180) & 289 (320) \\ 
\hline 
g [nm] & 145 (140) & 250 (280) \\ 
\hline 
h [nm] & 140 (120) & 150 (140) \\  
\hline
\hline
$|r|/|t|$ & (0.42/0.42) & (0.5/0.48)\\
\hline
\hline
$2\phi_{rt} [\degree]$ & $170\degree$ ($180\degree$) & $10\degree$ ($0\degree$) \\
\end{tabular} 
 \end{ruledtabular}
 \end{table}

\begin{figure}
\includegraphics[width=0.45\textwidth]{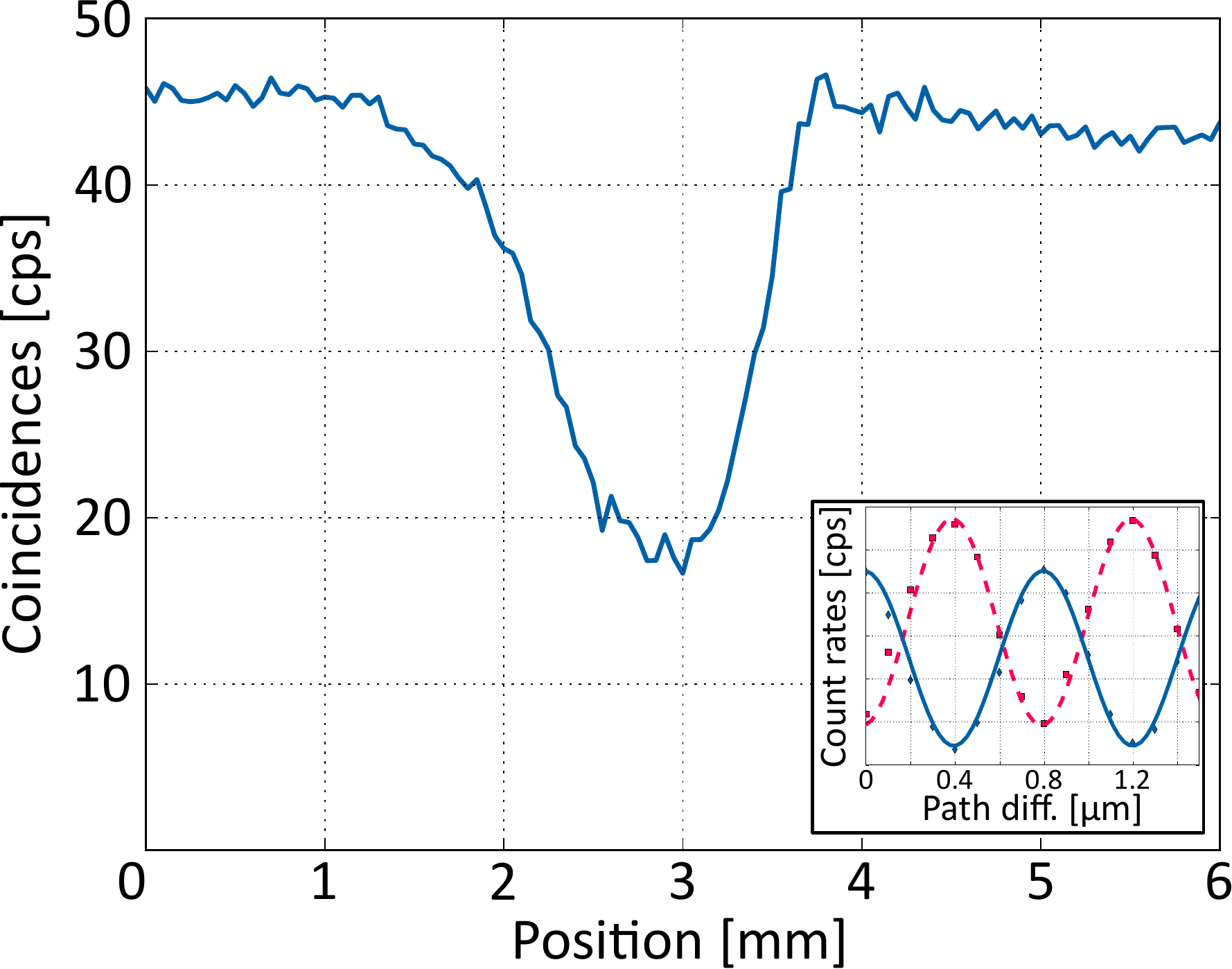}%
\caption{\label{fig3} Observation of a plasmonic Hong-Ou-Mandel coalescence effect with freely propagating single SPPs on sample I. The plot displays the coincidence count rates with respect to the delay $\delta_{\mathrm{HOM}}$ between both particles. The contrast of the dip is approximately $61\% \pm 2 \%$, further than the quantum limit at 50\% . The inset displays short interferograms of the classical fringes recorded by SPCMs A and B. Similarly to the lossless configuration, the observed sine waves are in phase opposition.}
\end{figure}

Figure \ref{fig3} is a plot of the coincidence rate with respect to the path difference between both arms when sample I is used. The inset is a plot of the sinusoidal fringes obtained at the outputs of the beamsplitter when illuminating with a laser at 806 nm. It is seen that the fringes are in phase opposition, confirming the $\frac{\pi}{2}$ phase-shift between $r$ and $t$. The plot displays a HOM-like dip, with a 61\% contrast, unambiguously in the quantum regime beyond the 50\% limit \cite{Ou:89}. This result is analog to the coalescence effect observed in two-photon quantum interference. This confirms the bosonic behavior of single plasmon, here achieved with freely-propagating, non-guided SPPs on a gold surface.

\begin{figure}
\includegraphics[width=0.45\textwidth]{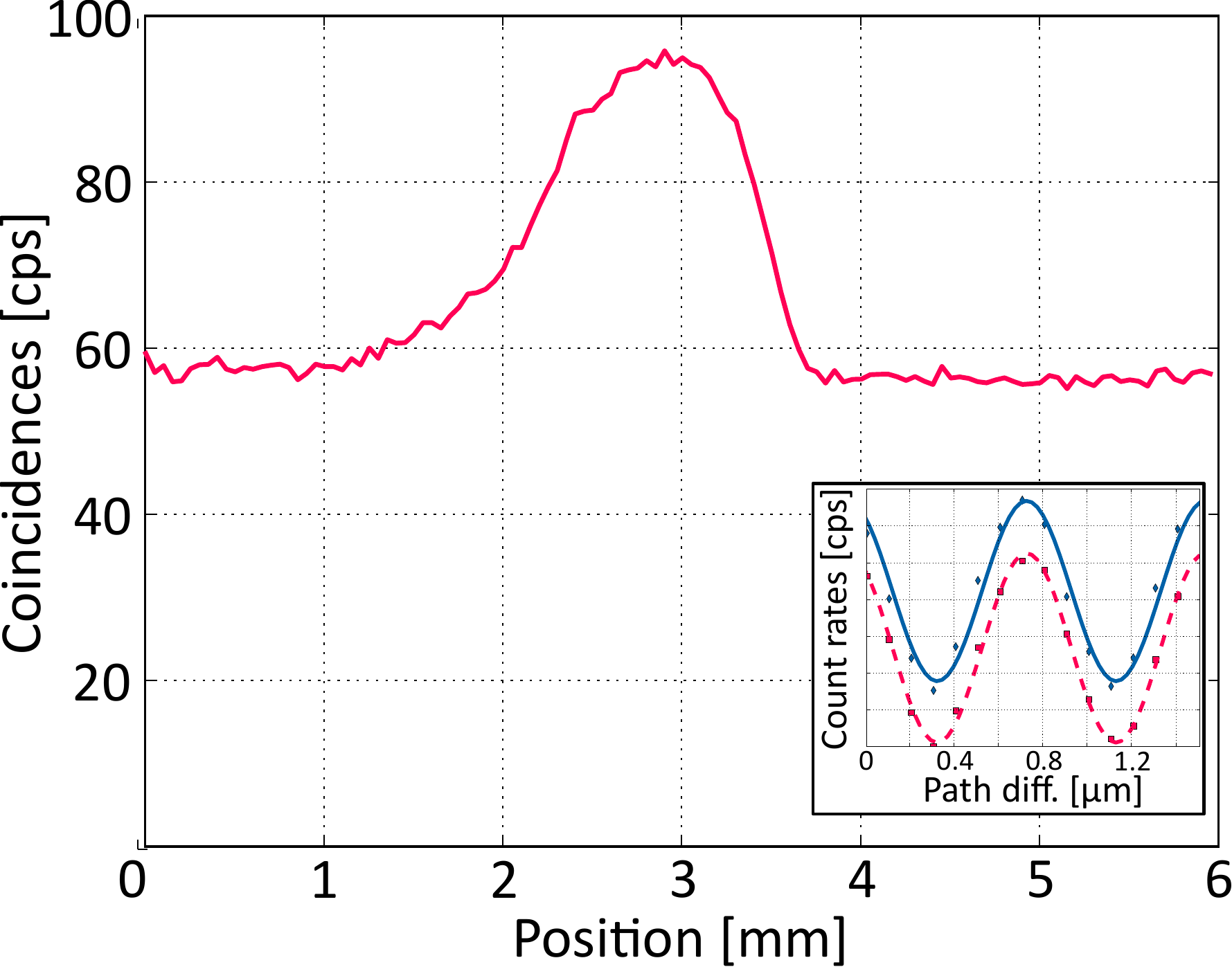}%
\caption{\label{fig4}Observation of a plasmonic Hong-Ou-Mandel peak when using sample II. We plotted the coincidence count rate between SPCMs A and B for a varying delay between the particles interfering on the SPBS.  The contrast of the peak is $72\% \pm 2 \%$. Here, the SPBS coefficients have been chosen so that the classical sine fringes are in-phase (see inset). This observation can be interpreted as an anti-coalescence effect.}
\end{figure}

We then move to the next sample, beamsplitter II. Inset of Fig. \ref{fig4} shows that classical fringes at the output are in-phase. In this case, orthogonality is not preserved between output modes of the SPBS. Two-particle quantum interference experiment is now characterized by a HOM-peak, an increase in coincidence rate with respect to the classical case. The contrast is around 72\%, once again in a quantum regime. The peak illustrates that when combining on this beamsplitter, SPPs tends to emerge from two different outputs. This anti-coalescence effect highlights the fundamental role of the SPBS in quantum interference.

In summary, we have observed quantum interference between two non-guided undistinguishable SPPs on lossy beamsplitters in a plasmonic version of the Hong-Ou-Mandel experiments. The single SPPs were excited by single photons delivered by a pair source. By changing the geometrical parameters of the beamsplitters, we could modify the relative phase between the reflection and transmission coefficients. Depending on this phase, we could observe either the expected HOM-dip illustrating coalescence between bosonic particles, or a more exotic HOM-peak, i.e. anti-coalescence resulting from the influence of losses and destructive interference between lossy path amplitudes.

\subsection{Methods}

\textbf{Detection method.} All the photons in these experiments were sent to SPCMs, which deliver transistor-transistor logic pulses. SPCMs A and B are Perkin-Elmer modules (SPCM AQRH-14). To count the correlations between the SPCMs A and B pulses, we used a PXI Express system from National Instruments (NI). The NI system is composed of a PXIe-1073 chassis on which NI FlexRIO materials are plugged : a field programmable gate array (FPGA) chip (NI PXIe-7961R) and an adapter module at 100 MHz (NI 6581). The FPGA technology allows changing the setting of the acquisition by simply programming the FPGA chip to whatever set of experiments we want to conduct. A rising edge from SPCM A or B triggers the detection of another rising edge respectively on channel B or A at specific delays. Counting rates and coincidences between channels A and B are registered. The resolution of the detection system is mainly ruled by the acquisition board frequency clock at 100 MHz, which corresponds to a time resolution of 10 ns. 

\textbf{Photon pair source.} The photon pairs source is based on parametric down-conversion. A potassium titanyl phosphate crystal (PPKTP crystal from Raicol) crystal is pumped at 403 nm by a tunable laser diode (Toptica). It delivers a 38 mW powered-beam, focused in the crystal by a 300 mm focal length planoconvex lens. The waist in the crystal is estimated to be 60 $\mu$m. The crystal generates pairs of orthogonally polarized photons at 806 nm. The waist in the crystal is conjugated to infinity with a 100-mm focal-length plano-convex lens, and the red photons emerging from the crystal are separated in polarization by a polarizing beamsplitter (PBS) cube (Fichou Optics). We remove the remaining pumping signal with a 1-nm-spanned interference filter (IF) from AHF (FF01-810/10).

\textbf{Coupling to the platform and collection of the output photons.} The photons in modes $\phi_1$ and $\phi_2$ are coupled to polarization-maintaining monomode fibers (P1-780PM-FC) via collimators (F220FC-780, Thorlabs). Each photon is outcoupled via Long Working Distance M Plan Semi-Apochromat microscope objectives (LMPLFLN-20X BD, Olympus) and sent to two different inputs of a PBS (Fichou Optics) with orthogonal polarizations. They leave the cube by the same output port and were focused with a 10X microscope objective (Olympus) on the plasmonic sample. The plasmonic sample is mounted on a solid immersion lens. The surface plasmons propagating on the chip leave the sample by two orthogonal output slits. The conversion of the SPPs back to photons via the slits leads to two different directions of propagation in free space. The photons from the output ports are collected from the rear side of the sample using mirrors and a 75-mm focal-length lens for each output. The output modes are then conjugated to multimode fibers via a 10X microscope objective (Olympus), which are connected to the SPCMs.

\textbf{Plasmonic platform sample fabrication.} We deposited 300-nm-thick gold films on clean glass substrates by e-beam evaporation (ME300 Plassys system) at a pressure of $2.10^{-6}$ mbar and at a rate of 0.5 nm/s. The rms roughness is 1 nm. The films were then loaded in a crossbeam Zeiss Auriga system and milled by a focused ion beam at low current (20 pA), except for the large slits used to decouple plasmons for propagating light that were milled at 600 pA.

\subsection{Acknowledgments}

We acknowledge the contribution of P. Lalanne and of J.-C. Rodier for the design of the plasmonic chip. We also acknowledge F. Cadiz, and N. Schilder for their help in the beginning of this study, as well as L. Jacubowiez, A. Browaeys, and P. Grangier for fruitful discussions. 

The research was supported by a DGA-MRIS (Direction Générale de l’Armement– Mission Recherche et Innovation Scientifique) scholarship, by RTRA (Réseau Thématique de Recherche Avancée) Triangle de la Physique, by the SAFRAN-IOGS chair on Ultimate Photonics and by a public grant overseen by the French National Research Agency (ANR) as part of the « Investissements d’Avenir » program (reference: ANR-10-LABX-0035, Labex NanoSaclay). J.-J.G. acknowledges the support of Institut Universitaire de France. 

\subsection{Author contributions}

G.M., F.M., and J.-J.G. initiated the project. J.-P.H. designed the chip, which was fabricated by E.R., E.D. and T.W.E. and characterized by A.B. and M.-C.D. M.-C.D. built the setup. Quantum experiments and data analysis were performed by B.V. under the supervision of F.M., and J.-J. G. B.V. , F.M. and J.-J.G. wrote the paper and discussed the results.

\subsection{Competing financial interests}

The authors declare no competing financial interests.

%

\end{document}